\input{aipcheck}

\documentclass[
    ,final            % use final for the camera ready runs
  ,numberedheadings % uncomment this option for numbered sections
  ,cmfonts                 % add further options here if necessary
  ]
  {aipproc}

\layoutstyle{6x9}

\begin{document}
\title{Proto-Neutron Star Winds, Magnetar Birth, and Gamma-Ray Bursts}
\classification{97.60.Bw, 97.60.Gb; 97.10.Me}
\keywords      {neutron stars, stellar winds, supernovae, gamma ray bursts, magnetic fields}
\author{Brian D. Metzger}{address = {Astronomy Department and Theoretical Astrophysics Center, 601 Campbell Hall, Berkeley, CA 94720; bmetzger@astro.berkeley.edu, eliot@astro.berkeley.edu},altaddress={Department of Physics, 366 LeConte Hall, University of California, Berkeley, CA 94720} % additional visiting address
}
\author{Todd A. Thompson}{ address = {Department of Astrophysical Sciences, Peyton Hall-Ivy Lane, Princeton University, Princeton, NJ 08544; thomp@astro.princeton.edu}}
\author{Eliot Quataert}{ address = {Astronomy Department and Theoretical Astrophysics Center, 601 Campbell Hall, Berkeley, CA 94720; bmetzger@astro.berkeley.edu, eliot@astro.berkeley.edu}}
  
%address={<common address for author2 and author3>}

\begin{abstract}
We begin by reviewing the theory of thermal, neutrino-driven proto-neutron star (PNS) winds.  Including the effects of magnetic fields and rotation, we then derive the mass and energy loss from magnetically-driven PNS winds for both relativistic and non-relativistic outflows, including important multi-dimensional considerations.  With these simple analytic scalings we argue that proto-magnetars born with $\sim$ millisecond rotation periods produce relativistic winds just a few seconds after core collapse with luminosities, timescales, mass-loading, and internal shock efficiencies favorable for producing long-duration gamma-ray bursts.                
\end{abstract}
\maketitle
%%%%%%%%%%%%%%%%%%%%%%%%%%%%%%%%%%%%%%%%%%%%
%% MAINMATTER
%%%%%%%%%%%%%%%%%%%%%%%%%%%%%%%%%%%%%%%%%%%%
\section{Neutrino-Driven PNS Winds}
\label{section:neutrino-driven}
  After a successful core-collapse supernova (SN), a hot proto-neutron star (PNS) cools and deleptonizes, releasing the majority of its gravitational binding energy ($\sim 3 \times 10^{53}$ ergs) in neutrinos.  With initial core temperature $T > 10$ MeV, a PNS is born optically-thick to neutrinos of all flavors because the relevant neutrino-matter cross sections scale as $\sigma_{\nu n} \propto \epsilon_{\nu}^{2} \propto T^{2}$, where $\epsilon_{\nu}$ is a typical neutrino energy.  Indeed, because neutrinos are trapped, a PNS's neutrino luminosity $L_{\nu}$ remains substantial and quasi-thermal for a time after bounce $\tau_{{\rm KH}} \sim 10-100$ s, as roughly verified by the 19 neutrinos detected from SN1987A 20 years ago [1],[2].  Although this Kelvin-Helmholtz (KH) cooling epoch is short compared to the time required for the shock, once successful and moving outward at $\sim 10^{4}$ km/s, to traverse the progenitor stellar mantle, $\tau_{{\rm KH}}$ is still significantly longer than the time over which the initial explosion must be successful.  While the specific shock launching mechanism is presently unknown, it must occur in a time $t <$ 1 s $\ll \tau_{{\rm KH}}$ after bounce for the PNS to avoid accreting too much matter.

Thus, even after the SN shock has cleared a cavity of relatively low density material around the PNS, $L_{\nu}$ remains substantial.  Detailed PNS cooling calculations [3] show that the electron neutrino(antineutrino) luminosity $L_{\nu_{e}(\bar{\nu}_{e})}$ is $ \sim 10^{52}$ erg/s at $t \sim 1$ s and declines as $\propto t^{-1}$ until $t \simeq \tau_{\rm KH}$, after which $L_{\nu_{e}(\bar{\nu}_{e})}$ decreases exponentially as the PNS becomes optically thin.  This persistent neutrino flux $F_{\nu_{e}(\bar{\nu}_{e})}$ continues to heat the PNS atmosphere, primarily through electron neutrino(antineutrino) absorption on nuclei ($\nu_{e}+n\rightarrow p+e^-$ and $\bar{\nu}_{e}+p\rightarrow n+e^+$).  Because the inverse, pair capture rates dominate the cooling, which declines rapidly with temperature ($\dot{q}^{-} \propto T^{6}$) and hence with spherical radius $r$, a region of significant net positive heating ($\dot{q} \equiv \dot{q}^{+} - \dot{q}^{-} >0$) develops above the neutrinosphere radius $R_{\nu}$.  This heating drives mass-loss from the PNS in the form of a thermally-driven wind [4].  To estimate the dependence of the resultant mass-loss rate ($\dot{M}_{\rm th}$) on the PNS properties explicitly, consider that in steady state the change in gravitational potential required for a unit mass element to escape the PNS ($GM/R_{\nu}$) must be provided by the total heating it receives accelerating outwards from the PNS surface:
\begin{equation} 
{\frac {GM}{R_{\nu}} \approx \int_{R_{\nu}}^{\infty}\dot{q} \frac{dr}{v_{r}}},
\label{bernoulli}
\end{equation}
where $M$ is the PNS mass, $v_{r}$ is the outward wind velocity, and $\dot{q}$ is per unit mass.  Because $\dot{q}$ is quickly dominated by heating from neutrino absorption, which scales as $\dot{q}^{+} \propto F_{\nu_{e}}\sigma_{n\nu} \propto  L_{\nu_{e}}\epsilon_{\nu_{e}}^{2}/4\pi r^{2}$, we see that equation (\ref{bernoulli}) implies that
\begin{equation} 
{\frac {GM}{R_{\nu}} \propto \frac{L_{\nu_{e}}\epsilon_{\nu_{e}}^{2}}{\dot{M}_{\rm th}}\int_{R_{\nu}}^{\infty}\rho dr \approx  \frac{L_{\nu_{e}}\epsilon_{\nu_{e}}^{2}}{\dot{M}_{\rm th}}\rho_{\nu}H_{\nu}},
\label{bernoulli2}
\end{equation}
where we have used $\dot{M}_{\rm th} = 4\pi\rho r^{2}v_{r}$ for a spherical wind, $\rho$ is the mass density, $H$ is the PNS's density scale height, $\epsilon_{\nu_{e}}$ crudely defines a mean electron neutrino or antineutrino energy, and a subscript ``$\nu$'' denotes evaluation near $R_{\nu}$.  Neglecting rotational support and assuming that the thermal pressure $P$ is dominated by photons and relativistic pairs (which also becomes an excellent approximation as the density plummets abruptly above the PNS surface), we have that $H_{\nu} \sim P_{\nu}/\rho_{\nu} g_{\nu} \propto T_{\nu}^{4}R_{\nu}^{2}/M\rho_{\nu}$, where $g_{\nu} \propto M/R_{\nu}^{2}$ is the PNS surface gravity and $T_{\nu} \propto (L_{\nu_{e}}\epsilon_{\nu_{e}}^{2}/R_{\nu}^{2})^{1/6}$ is the PNS surface temperature.  $T_{\nu}$ is set by the balance between heating and cooling at the PNS surface ($T_{\nu}^{6} \propto \dot{q}^{-} = \dot{q}^{+} \propto L_{\nu_{e}}\epsilon_{\nu_{e}}^{2}/R_{\nu}^{2}$).  Inserting these results into equation (\ref{bernoulli2}) and including the correct normalization from the relevant weak cross sections, one finds the expression for $\dot{M}_{\rm th}$ first obtained by ref [4]:
\begin{equation}
 {\dot{M}_{\rm th} \approx 10^{-4}L_{52}^{5/3}\,\epsilon_{10}^{10/3}M_{1.4}^{-2}R_{10}^{5/3}\,M_{\odot}{\rm /s}},
\label{mdot} 
\end{equation}
where $L_{52} \equiv L_{\nu_{e}}\times10^{52}$ erg/s, $\epsilon_{10} \equiv 10\epsilon_{\nu_{e}}$MeV, $R_{\nu} \equiv 10 R_{10}$ km, and $M\equiv1.4M_{1.4} M_{\odot}$.  

Endowed with an enormous gravitational binding energy and a means, through this neutrino-driven outflow, for communicating a fraction of this energy to the outgoing shock, a newly-born PNS seems capable of affecting the properties of the SN that we observe.  However, a purely thermal, neutrino-driven PNS wind is only accelerated to an asymptotic speed of order the surface sound speed: $v^{\infty}_{\rm th} \sim c_{s,\nu} \approx \sqrt{2kT_{\nu}/m_{p}} \approx 0.1 L_{52}^{1/12}\epsilon_{10}^{1/6}R_{10}^{-1/6}$ c.  Thus, the efficiency $\eta$ relating wind power $\dot{E}_{\rm th} \approx \dot{M}_{\rm th}(v^{\infty}_{\rm th})^{2}/2$ to total neutrino luminosity ($L_{\nu} \sim 6 L_{\nu_{e}}$) is quite low:
\begin{equation}
\eta \equiv \frac{\dot{E}_{\rm th}}{L_{\nu}} \sim 10^{-5}L_{52}^{5/6}\epsilon_{10}^{11/3}R_{10}^{4/3}M_{1.4}^{-2}.
\end{equation}
In particular, although neutrino energy deposited in a similar manner may be responsible for initiating the SN explosion itself at early times (i.e., the neutrino SN mechanism [5]), $\eta$ drops rapidly as the PNS cools.  Quasi-spherical winds of this type are therefore not expected to affect the SN's nucleosynthesis or morphology (although the wind itself is considered a promising $r$-process source [4]).
\section{Magnetically-Driven PNS Winds}
Some PNSs may possess a more readily extractable form of energy in rotation.  A PNS born with a period $P = P_{\rm ms}$ ms is endowed with a rotational energy $E_{\rm rot} \simeq 2\times 10^{52}P_{\rm ms}^{-2}R_{10}^{2}M_{1.4}\,{\rm ergs}$, which, for $P < 4$ ms, exceeds the energy of a typical SN shock ($\sim 10^{51}$ ergs).  Given a mass loss rate $\dot{M}$ and torquing lever arm $\omega_{\tau}$, a wind extracts angular momentum $J$ from the PNS at a rate $\dot{J} \simeq \Omega \omega_{\tau}^{2}\dot{M}$, where $\Omega = 2\pi/P$ is the PNS rotation rate.   With the PNS's radius $R_{\nu}$ as a lever arm and the modest thermally-driven mass-loss rate given by equation (\ref{mdot}), the timescale for removal of the PNS's rotational energy, $\tau_{J} \equiv J/\dot{J} \sim MR_{\nu}^{2}/\dot{M}\omega_{\tau}^{2}\sim M/\dot{M}_{\rm th}$, is much longer than $\tau_{{\rm KH}}$.  However, if the PNS is rapidly rotating and possesses a dynamically-important poloidal magnetic field $B_{p}$ (through either flux-freezing or generated via dynamo action [6]), then both $\dot{M}$ and $\omega_{\tau}$ can be substantially increased; this reduces $\tau_{J}$, allowing efficient extraction of $E_{\rm rot}$.

For magnetized winds $\omega_{\tau}$ is the Alfv\'{e}n radius $\omega_{A}$, defined as the cylindrical radius where $\rho v_{r}^{2}/2$ first exceeds $B_{p}^{2}/8\pi$ [7].  The magnetosphere of a PNS is most likely dominated by its dipole component, with a total (positive-definite) surface magnetic flux given by $\Phi_{\rm B} = 2\pi B_{\nu}R_{\nu}^{2}$, where $B_{\nu}$ is the polar surface field.  To estimate $\omega_{A}$ for magnetized PNS outflows recognize that mass and angular momentum are primarily extracted from a PNS along open magnetic flux.  For an axisymmetric dipole rotator this represents only a fraction $\approx 2(\pi\theta_{\rm LCFL}^{2})/4\pi \simeq R_{\nu}/2\omega_{\rm Y}$ of $\Phi_{\rm B}$, where $\theta_{\rm LCFL} \approx \sqrt{R_{\nu}/\omega_{\rm Y}}$ is the latitude (measured from the pole) at the PNS surface of the last closed field line (LCFL), $\omega_{\rm Y}$ is the radius where the LCFL intersects the equator (the ``Y point''), and we have assumed that $\omega_{\rm Y} \gg R_{\nu}$ ($\theta_{\rm LCFL} \ll 1$).   Plasma necessarily threads a PNS's closed magnetosphere and cannot be forced to corotate superluminally; thus $\omega_{\rm Y}$ cannot exceed the light cylinder radius $\omega_{\rm L} \equiv c/\Omega = 48P_{\rm ms}$ km, making it useful to write the PNS magnetosphere's total open magnetic flux as $\Phi_{\rm B,open} \approx \pi B_{\nu}R_{\nu}^{2}(R_{\nu}/\omega_{\rm L})(\omega_{\rm Y}/\omega_{\rm L})^{-1}$.  Now, the overall \textit{latitudinal structure} of a PNS magnetosphere (i.e., the allocation of open and closed magnetic flux, and the value of $\omega_{\rm Y}/\omega_{\rm L}$) is primarily dominated by the dipolar closed zone.  However, recent numerical simulations [8] show that where the field is open it behaves as a ``split monopole''.  In this case the poloidal field scales as $B_{p} \sim \Phi_{\rm B,open}/r^{2} \approx 0.2B_{\nu}P_{\rm ms}^{-1}R_{10}(\omega_{\rm Y}/\omega_{\rm L})^{-1}(R_{\nu}/r)^{2}$, rather than the dipole scaling $\propto (R_{\nu}/r)^{3}$.  The constant of proportionality is chosen to assure that $B_{p}(R_{\nu}) \rightarrow B_{\nu}$ in the limit of vanishing closed zone ($\omega_{\rm L},\omega_{\rm Y} \rightarrow R_{\nu}$) and is in agreement with numerical results (see eq.~[28] of ref [8]).
%\begin{center}
%{\bf Non-Relativistic Winds and Asymmetric Supernovae}
%\end{center}
%\subsection{Non-Relativistic Winds}
\subsection{Non-Relativistic Winds and Asymmetric Supernovae}
Non-relativistic (NR) magnetically-driven winds reach an equipartition between kinetic and magnetic energy outside $\omega_{A}$ such that the kinetic energy flux at $\omega_{A}$ ($\dot{M}v_{r}(\omega_{A})^{2}/2$) carries a sizeable fraction of the rotational energy loss extracted by the wind's surface torque $\dot{E}_{\rm rot} = \dot{J}\Omega = \dot{M}\Omega^{2}\omega_{A}^{2}$; thus, we have that $v_{r}(\omega_{A}) \sim \Omega \omega_{A}$.  Combining this with the modified monopole scaling for $B_{p}$ motivated above and mass conservation $\dot{M}_{\Omega} \equiv \rho r^{2}v_{r}$ ($\dot{M}_{\Omega}$ is the mass flux \textit{per solid angle}) we find that:
\begin{equation}
\omega_{\rm A}/R_{\nu} \simeq B_{15}^{2/3}P_{\rm ms}^{-2/3}\dot{M}_{\Omega,-4}^{-1/3}R_{10}^{4/3}(\omega_{\rm Y}/\omega_{\rm L})^{-1},
\label{ra}
\end{equation}
where $\dot{M}_{\Omega} \equiv \dot{M}_{\Omega,-4}\times 10^{-4} M_{\odot}$s$^{-1}$sr$^{-1}$, $B_{\nu} \equiv B_{15}\times 10^{15}$ G, and we have concentrated on the open magnetic flux that emerges nearest the closed zone (polar latitude $\approx \theta_{\rm LCFL}$) and which thereby dominates the spin-down torque.  

From equation (\ref{ra}) we see that winds from rapidly rotating PNSs with surface magnetic fields typical of Galactic ``magnetars'' ($B_{\nu} \sim 10^{14}-10^{15}$ G) possess enhanced lever arms for extracting rotational energy [9].  Furthermore, their total outflow power $\dot{E}_{\rm mag}^{\rm NR} \approx \dot{E}_{\rm rot} \approx 2\pi\theta_{\rm LCFL}^{2}\dot{M}_{\Omega}\Omega^{2}\omega_{A}^{2} \approx 10^{49}B_{15}^{4/3}P_{\rm ms}^{-13/3}\dot{M}_{\Omega,-4}^{1/3}R_{10}^{17/3}(\omega_{\rm Y}/\omega_{\rm L})^{-3}$ ergs/s dominates thermal acceleration ($\dot{E}_{\rm mag}^{\rm NR} > \dot{E}_{\rm th}$) for $B_{15} > 0.4 P_{\rm ms}^{13/4}L_{52}^{23/24}\epsilon_{10}^{23/12}R_{10}^{-11/3}M_{1.4}^{-1}(\omega_{\rm Y}/\omega_{\rm L})^{9/4}$.  This condition becomes easier to satisfy as the PNS cools, allowing magnetized winds to dominate later stages of the KH epoch for PNSs with even relatively modest $B_{\nu}$ and $\Omega$.  NR magnetically-driven winds, in addition to being more powerful than spherical, thermally-driven outflows, are efficiently hoop-stress collimated along the PNS rotation axis [8].  The power they deposit along the poles may produce asymmetry in SN ejecta distinct from the shock-launching process itself.

Strong magnetic fields and rapid rotation can also increase the outflow's power through enhanced mass-loss because $\dot{E}_{\rm mag}^{\rm NR} \propto \dot{M}_{\Omega}^{1/3}$.  When the PNS's hydrostatic atmosphere is forced to co-rotate to the outflow's sonic radius $\omega_{\rm s} = (GM\sin[\theta_{\rm LCFL}]/\Omega^{2})^{1/3}$ then $\dot{M}_{\Omega}$ is enhanced by a factor $\phi_{\rm cf} \sim \exp[(v_{\phi,\nu}/c_{s,\nu})^{2}]$ over $\dot{M}_{\rm th}/4\pi$ due to centrifugal (``cf'') slinging [9], where $v_{\phi,\nu} \approx R_{\nu}\Omega\sin[\theta_{\rm LCFL}] \approx R_{\nu}\Omega\sqrt{R_{\nu}/\omega_{\rm Y}}$ is the PNS rotation speed at the base of the open flux.  Using our estimate for $c_{s,\nu}$ from $\S$ 1, we see that enhanced mass loss becomes important for $P_{\rm ms} < P_{\rm cf,ms} \equiv L_{52}^{-1/18}\epsilon_{10}^{-1/9}R_{10}^{10/9}(\omega_{\rm Y}/\omega_{\rm L})^{-1/3}$ (i.e., only for PNSs with considerable rotational energy $E_{\rm rot} > 10^{52}$ ergs).  Fully enhanced mass loss ($\dot{M}_{\Omega} = \dot{M}_{\rm th}\phi_{\rm cf}/4\pi$) requires $\omega_{A} > \omega_{\rm s}$, which in turn requires that $B_{15} > B_{\rm cf,15} \equiv P_{\rm ms}^{7/4}R_{10}^{-13/4}\dot{M}_{\Omega,-4}^{1/2}M_{1.4}^{1/2}(\omega_{\rm Y}/\omega_{\rm L})^{5/4} \simeq 0.3 P_{\rm ms}^{7/4}L_{52}^{5/6}\,\epsilon_{10}^{5/3}M_{1.4}^{-1/2}R_{10}^{-29/12}\exp[0.5(P/P_{\rm cf})^{-3}](\omega_{\rm Y}/\omega_{\rm L})^{5/4}$, where we have taken $\dot{M}_{\rm th}$ from $\S$ 1.  For cases with $B_{\nu} < B_{\rm cf}$ but $P < P_{\rm cf}$, $\dot{M}_{\Omega}$ lies somewhere between $\dot{M}_{\rm th}/4\pi$ and $\phi_{\rm cf}\dot{M}_{\rm th}/4\pi$ (see [10] for numerical results).  Millisecond proto-magnetars generally attain $\phi_{\rm cf}$, except perhaps at early times when the PNS is quite hot.
%\appendix
%\begin{center}
%{\bf Relativistic Winds and Gamma-Ray Bursts}
%\end{center}
\subsection{Relativistic Winds and Gamma-Ray Bursts}
As the PNS cools, eventually $\omega_{A}\rightarrow \omega_{\rm L}$ and the PNS outflow becomes relativistic (REL).  This transition occurs after $\tau_{\rm KH}$ for most PNSs (they become pulsars), but rapidly rotating proto-magnetar winds become relativistic during the KH epoch itself.  Similar to normal pulsars, PNSs of this type lose energy at the force-free, ``vacuum dipole'' rate: $\dot{E}^{\rm REL}_{\rm mag} \approx 6\times 10^{49}B_{15}^{2}P_{\rm ms}^{-4}R_{10}^{6}(\omega_{\rm Y}/\omega_{\rm L})^{-2}\,{\rm ergs/s}$ (again modulo corrections for excess open magnetic flux $\dot{E}_{\rm mag}^{\rm REL} \propto \Phi_{\rm B, open}^{2} \propto (\omega_{\rm Y}/\omega_{\rm L})^{-2}$ [8]), which gives a familiar spin-down timescale $\tau_{\rm J} = E_{\rm rot}/\dot{E}^{\rm REL}_{\rm mag} \approx 300 B_{15}^{-2}P_{\rm ms}^{2}R_{10}^{-4}M_{1.4}(\omega_{\rm Y}/\omega_{\rm L})^{2}$ s.  On the other hand, the mass loading on a PNS's open magnetic flux is set by neutrino heating, a process totally different from the way that matter is extracted from a normal pulsar's surface.  In fact, a proto-magnetar outflow's energy-to-mass ratio $\sigma$ is given by
\begin{equation}
\sigma \approx \frac{\dot{E}^{\rm REL}_{\rm mag}}{2\pi\theta_{\rm LCFL}^{2}\dot{M}_{\Omega}c^{2}}  \approx 3B_{15}^{2}P_{\rm ms}^{-3}L_{52}^{-5/3}\epsilon_{10}^{-10/3}R_{10}^{10/3}M_{1.4}^{2}\exp\left[-\left(\frac{P}{P_{\rm cf}}\right)^{-3}\right]\left(\frac{\omega_{\rm Y}}{\omega_{\rm L}}\right)^{-1}
\label{sigma}
\end{equation}
From equation (\ref{sigma}) we see that because a PNS's mass-loss rate drops so precipitously as it cools, $\sigma \propto L_{\nu_{e}}^{-5/3}\epsilon_{\nu_{e}}^{-10/3}$ rises rapidly with time, easily reaching $\sim 10-1000$ during the KH epoch for typical magnetar parameters [9],[10].  Detailed evolution calculations indicate that $E_{\rm rot}$ is extracted roughly uniformly in log($\sigma$) [10].  

To conclude with a concrete example, consider a proto-magnetar with $B_{\nu} = 10^{16}$ G and $P_{\rm ms} = 3$ at $t = 10$ seconds after core collapse.  From the cooling calculations of ref [3] we have $L_{52}(10$ s$) \approx 0.1$ and $\epsilon_{10}(10$ s$) \approx 1$ (see Figs. [14] and [18]) and so, under the conservative estimate that $\omega_{\rm Y} = \omega_{\rm L}$, equation (\ref{sigma}) gives $\sigma \approx 500$.  Because $\sigma$ represents the potential Lorentz factor of the outflow (assuming efficient conversion of magnetic to kinetic energy), we observe that millisecond proto-magnetar birth provides the right mass-loading to explain gamma-ray bursts (GRBs).  Further, the power at $t=10$ s is still $\dot{E}^{\rm REL}_{\rm mag} \approx 10^{50}$ erg/s with a spin-down time $\tau_{\rm J} \approx 30$ s, both reasonable values to explain typical luminosities and durations, respectively, of long-duration GRBs.  Lastly, because $\sigma$ rises so rapidly with time as the PNS cools, in the context of GRB internal shock models a cooling proto-magnetar outflow's kinetic-to-$\gamma$-ray efficiency can be quite high; our calculations indicate that values of $10-50 \%$ are plausible.  We conclude that magnetar birth accompanied by rapid rotation (but requiring less angular momentum than collapsar models) represents a viable long-duration GRB central engine.

\end{document}